\documentclass[%
 reprint,
 amsmath,amssymb,
prl,
]{revtex4-2}

\usepackage{graphicx}
\usepackage{dcolumn}
\usepackage{bm}

\newcommand{\veb}[1]{\ifcat\noexpand#1\relax\bm{#1}\else\mathbf{#1}\fi}
\newcommand{\teb}[1]{\ifcat\noexpand#1\relax\bm{#1}\else\mathbf{#1}\fi}

\newcommand{\spatialaverage}[1]{\left\langle{#1}\right\rangle}

\newcommand{\Weissenberg}{\mathit{Wi}}

\begin{document}

\preprint{APS/123-QED}

\title{Comment on ``Kinematic and Rheological Equivalence of Steady Shearing and Planar Extensional Flows''}

\author{Fabian Hillebrand}%
 \email{fabian.hillebrand@oist.jp}
\affiliation{%
Okinawa Institute of Science and Technology Graduate University, 1919-1 Tancha, Onna, Okinawa 904-0495, Japan
}%

\date{\today}

\maketitle


In Ref.\  \cite{king2026kinematicandrheologicalequivalence}, the authors claim the `kinematic' and `rheological' equivalence between simple shear and planar extension.
They state that their result holds ``for any materially objective constitutive equation.''
The proposed relation is exact for single-mode constitutive models whose stress is coaxial with a single tensor obeying an evolution equation that depends linearly on the velocity gradient.
However, it fails for standard kinetic theories of dilute solutions, e.g.\ FENE dumbbells \cite{warner1972kinetictheoryfene}, and for the integral tube \cite{doi1986} or K-BKZ models \cite{bernstein1963astudy} of entangled solutions and melts.
Their result is a consequence of the employed closures, not of an underlying equivalence between the flows.
Its accuracy for real fluids cannot be inferred from such closed models.

\begin{figure}
    \centering
    \includegraphics[width=1.0\linewidth]{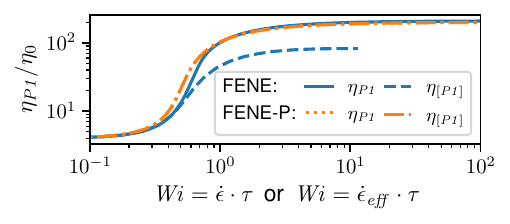}
    \caption{First planar extensional viscosity \(\eta_{\mathit{P1}}\) and proposed \(\eta_{[\mathit{P1}]}\) based on single-mode FENE dumbbells (blue, solid and dashed, respectively), using Brownian dynamics simulations, and FENE-P dumbbells (orange, dotted and dash-dotted, respectively), where the lines coincide, with finite extensibility \(b=100\) for both.}
    \label{fig:fene}
\end{figure}

Kinematically, simple shear and planar extension are distinct, which follows from Astarita's objective flow classification \cite{astarita1979objective}.
The authors' proposed transformation cannot remove the rotational rate \(\dot{\omega} = \dot{\gamma} / 2\) in simple shear and additionally introduces an effective shear rate \(\dot{\gamma}_\mathit{eff} = \dot{\gamma}N_1 / \sqrt{N_1^2+4\sigma_{12}^2}\), which must not couple to the principal stresses for their relation to hold.
Figure \ref{fig:fene} shows that the proposed rule systematically under-predicts the high-\(\Weissenberg\) plateau of \(\eta_{\mathit{P1}}\) for FENE dumbbells.
This is because, in simple shear, the viscoelastic stress \(\teb{\sigma}_p \sim \spatialaverage{\veb{r}\veb{f}(\veb{r})}\), with connector vector \(\veb{r}\) and spring force \(\veb{f}(\veb{r})\), does not commute with the second moment \(\spatialaverage{\veb{r}\veb{r}}\), whose evolution equation otherwise depends linearly on the velocity gradient. 
This failure occurs even for a single mode without solvent.
In contrast, the Peterlin closure in FENE-P expresses \(\teb{\sigma}_p\) as a function of \(\spatialaverage{\veb{r}\veb{r}}\), which forces them to commute and the relation holds.

Similar issues arise for integral models of entangled solutions and melts.
Here, the stress depends on the full strain history, whose principal axes differ from those of the stress, which breaks the rule for nonlinear damping.
Only the single-mode Lodge model, which has no damping and is equivalent to UCM, collapses to a single tensor without coupling \(\dot{\omega}\) or \(\dot{\gamma}_\mathit{eff}\).
The Doi--Edwards model with independent alignment approximation and a single relaxation time and the K-BKZ model with Wagner damping (damping exponent \(n=0.2\)) \cite{wagner1978nonlinearshaercreep} over-predict \(\eta_{\mathit{P1}}\) by up to \(26\%\) (\(\Weissenberg \approx 2.29\)) and \(74\%\) (\(\Weissenberg \approx 1.04\)), respectively.
In contrast, the Rolie-Poly model follows from a closure of the tube theory \cite{likhtman2003roliepoly}, and hence the authors' relation applies.
As a competing effect, multi-modality generally results in an under-prediction of \(\eta_{\mathit{P1}}\), since the critical extension rate is set by the viscosity-weighted rather than the longest relaxation time \cite{king2026newapproachmeasuringsteady}.
For example, the proposed transformation overestimates the critical extension rate by a factor of \(1.73\) for a good solvent and \(2.17\) for a \(\theta\)-solvent based on the Zimm model \cite{zimm1956dynamicsofpolymermolecules}, which is exactly a multi-mode UCM model.
Similarly, unless the solvent viscosity is subtracted as suggested by Ref.\  \cite{king2026newapproachmeasuringsteady}, it systematically shifts the critical extension rate by a factor \(1/(1-\beta)\) based on the Oldroyd-B model, with solvent viscosity ratio \(\beta = \eta_s/\eta_0\).

The authors suggest that their result augments the heuristic Cox--Merz rule \cite{cox1958correlation}.
However, the Cox--Merz rule is based on substantial experimental validation and its deviations are well characterized.
This is not the case for their result: because the competing effects may or may not cancel in entangled solutions and melts, the error cannot be assessed from shear data alone.
Therefore, the validity cannot be established without additional extensional data, which is precisely the measurement the rule was proposed to replace.
The only comparison with measured extensional data provided by the authors is for an LDPE melt, for which \(N_1\) was estimated from an empirical rule rather than measured, and \(\eta_{\mathit{P1}}\) stems from an entrance flow using a slit die rather than a controlled planar extensional flow.
The PIB data only compare shear data with a fitted model, where the result holds by construction.
Therefore, this comparison does not test the validity of their relation for this solution.

I acknowledge funding from the Japan Society for the Promotion of Science (JSPS, Grant No. 26KJ2101).
I further acknowledge the use of Claude Opus 5.5 and Sonnet 5.5 (Anthropic) to generate the code for the Brownian dynamics simulations and the integral model calculations, and to review and suggest edits to the manuscript.
I verified all code and results.
I thank Amy Q.\ Shen and Simon J.\ Haward for their suggestions, and Nicholas King and Gareth H.\ McKinley for the helpful correspondence.

\bibliography{apssamp}

\end{document}